\documentclass[aps,prb,floatfix,twocolumn]{revtex4}

\usepackage{sqrcaps}
\usepackage[T1]{fontenc}

\usepackage{ulem}

\usepackage{graphicx}
\usepackage{dcolumn}
\usepackage{bm}
\usepackage{epstopdf}

 \usepackage{lineno}
 \usepackage{amssymb}

\begin{document}

\title{ The contextuality loophole {\it is} fatal for the derivation of Bell inequalities: \\ Reply to a comment by I. Schmelzer }
 
 \author{Theodorus M. Nieuwenhuizen$^{1,2}$ and  Marian Kupczynski$^3$}

\address{
  $^1$ Institute for Theoretical Physics,  Science Park 904,  1090 GL Amsterdam, The Netherlands \\
$^2$  International Institute of Physics,   Av. Odilon Gomes de Lima, 1722,   59078-400 - Natal-RN, Brazil  \\
$^3$ 2 D\'epartement de l' Informatique, Universit\'e du Qu\'ebec en Outaouais (UQO), Case postale 1250, succursale Hull, Gatineau, Quebec, J8X 3X 7, Canada
}

\begin{abstract}
Ilya Schmelzer wrote recently: 
{\it Nieuwenhuizen argued that there exists some ``contextuality loophole'' in Bell's theorem. This claim in unjustified}. 
It is made clear that this arose from attaching a meaning to the title and the content of the paper different from the one intended by Nieuwenhuizen.
``Contextual loophole'' means only that if the supplementary parameters describing measuring instruments are correctly introduced, 
Bell and Bell-type inequalities may not be proven. It is also stressed that a hidden variable model suffers from a ``contextuality loophole'' if it tries to describe different 
sets of incompatible experiments using a unique probability space and a unique joint probability distribution.
\end{abstract}

 \maketitle


Ilya Schmelzer wrote recently: {\it Nieuwenhuizen argued that there exists some ``contextuality loophole'' in Bell's theorem} \cite{1}. {\it This claim in unjustified}. 
We shall point out that Schmelzer gave a meaning to the title and the content of the paper different form the one intended by Nieuwenhuizen.

First of all we must agree with Schmelzer that two paragraphs from  \cite{2}, which he cites, may lead to confusion if they are taken out of the context. 
In fact, local realism is used there as a possibility of explaining long range spin correlations in a local and a causal way. 
This is not what EPR and Bell meant by local realism. Of course one may find in literature various meanings given to the notion of local realism 
but the current consensus is that it should be understood as a counterfactual definiteness as it was used by Bell in 1964, 
next to signal speeds not exceeding the speed of light \cite{3}.
In the second part of his comment Schmelzer reproduces Bell's claim that the predetermination of outcomes in his model is derived and not postulated.
We disagree on that with Bell and Schmeltzer and we explain why the non-contextuality of the probabilistic model used in  \cite{3} is postulated and not derived. 
To make things clear, we explain below the -- in our view -- correct meaning of the term ``contextuality loophole''.

Bell's (1964) theorem  \cite{3} is based on a particular probabilistic model in which, for each EPR pair, 
spin projections in all directions are predetermined by a source and registered passively by measuring instruments. 
From this probabilistic model Bell rigorously deduces his inequalities thus there is no loophole in this proof nor in the theorem. 
Of course if the outcomes were predetermined there would be no reason to introduce additional parameters describing measuring instruments.

What Nieuwenhuizen stressed in his paper was that  this Bell 1964 model is totally unrealistic because it contains no hidden variables 
for the  measuring instruments and other relevant instruments, together called ``context'', so that it can not lead to any sound conclusion
about physics or Nature.

\newpage

We agree with Bell that, in his words,
{\it  if  one can predict in advance the result of measuring any chosen component of $\sigma_2$, by previously measuring the same component of $\sigma_1$, 
it  follows that the result of any such measurement must actually be predetermined} \cite{4}. 
(Here $\sigma_1$ and $\sigma_2$ are the spins of the particles measured by Alice and Bob respectively.)
The possibility of predicting an outcome without performing a measurement and its predetermination is then simply a tautology.

 The violation of all Bell-type inequalities under these conditions is impossible. 
 When Bell-type inequalities are violated in experiment, it means necessarily that the assumptions used to derive them  were not  general enough. 


In particular, the assumption that the outcomes of quantum measurements are predetermined, called in current literature {\it counterfactual definiteness} (CFD), 
is simply not valid in quantum physics. 
CFD is in conflict with the description of the measurements in quantum theory and with  Bohr's complementarity 
(the measurements of incompatible physical observables represented by non-commuting operators require mutually exclusive experimental contexts)  
see for example \cite{5,6,7a,7b,8,9,10}.  
A detailed discussion why Alice can not predict with certainty Bob's outcome in spin polarization experiments, and vice-versa,  may be found for example in  \cite{11,12,13}.

Quantum theory teaches us that outcomes of measurements are only created as the effect of a physical interaction of the measured system with 
the measuring instrument in a well defined experimental context \cite{6, 14,Opusculo,15,16,17}.
Therefore a hidden variable model which aims to explain details of quantum measurements  has to introduce context-dependent hidden variables,
 in particular  supplementary parameters describing measuring instruments during the measurement.  
 It is well known that if such contextual hidden variables are correctly introduced, Bell-type inequalities can not  be proven. 
 This is how one has to understand the title and the content of  Nieuwenhuizen's paper \cite{1}. 
 
 \newpage
 
 Bell claimed that if Bell-type inequalities were violated it would mean that causally local explanation of long range correlations in spin polarization 
 correlation experiments is impossible. Several authors clearly demonstrated that this claim is unjustified  \cite{2,15,16,17,18,19,20,21,22,23,24}
 and in \cite{2} this was termed the ``contextuality loophole''.

Since one can not predict individual outcomes in quantum theory and in particular not in spin polarization correlation experiments, 
it is obvious that, contrary to Schmelzer's claim, non-contextuality is a presupposition in Bell's (1964) analysis.

\newpage{}

\newpage {}

We hope to have explained clearly enough why and in what sense the contextuality loophole is fatal for derivation of Bell and Bell-type inequalities,
and that Schmelzer's rejection of it is based on his deviant interpretation.

 The contextuality loophole being a theoretical issue related to the handling of the experimental data,
 implies that it can not be closed in any experiment, so that when the data show
a violation of the Bell inequality, this is no more than an interesting but harmless property.


\vspace{1cm}


\end{document}